\documentclass[%
 reprint,
 amsmath,amssymb,
 aps,
prb,
]{revtex4-1}
\usepackage{graphicx}
\usepackage{dcolumn}
\usepackage{bm}
\begin{document}
\preprint{APS/PRB/Mass classification}
\title{Mass classification and manipulation of zero modes in one-dimensional Dirac systems}
\author{Yiming Pan}
\email{wilder.pan@gmail.com}
\author{Huaiqiang Wang}%
\author{D. Y. Xing}
\author{Baigeng Wang}%
\affiliation{National Laboratory of Solid State Microstructures and Department of Physics,
Nanjing University, Nanjing 210093, China}

\date{\today}
\begin{abstract}
We present a detailed mass classification of all possible zero-energy modes in one-dimensional Dirac systems. By introducing a linear mass term into the Dirac Hamiltonian, we find that the topologically protected zero-energy modes have the mass-momentum duality. Based on the duality, we classify three fundamental zero-energy modes in $2\times2$ subspaces respectively: solitons in sublattice subspace, Majorana zero modes in Nambu subspace, and magnetic zero-energy modes in spin subspace. Within the mechanism of mass competition, isolated zero-energy modes emerge by lifting KramerÕs degeneracy in the combined $8\times8$ inner space and its $4\times4$ subspaces. We also propose experimental methods of manipulating possible masses in Dirac Hamiltonians. Our classification scheme could easily be extended to 2D or 3D systems and applied to investigate topologically protected states in other fields.
\begin{description}
\item[PACS numbers] 
73.20.-r, 74.20.Rp,75.40.Cx,74.45.+c

\end{description}
\end{abstract}
\maketitle

\section{Introduction}
Classifications of topological phases are mainly based on symmetry analysis of single-particle Hamiltonians.\cite{ Schnyder08,1367-2630-12-6-065010} Three-dimensional topological insulators or superconductors exist in five out of ten symmetry classes, which are classified in terms of the presence or absence of time-reversal symmetry (TRS), particle-hole symmetry (PHS), and sublattice symmetry (SLS) within the random matrix theory. \cite{ Schnyder08}Topological nontrivial phases support topologically protected states at the boundary of bulk systems, such as zero-energy modes in one-dimensional (1D) systems, edge states in two-dimensional (2D) systems, and surface states in three-dimensional (3D) systems.\cite{Qi11,Hasan10} Recently, Ryu \textit{et al}. proposed a complete classification of all possible masses, 36 in total\cite{ Ryu09,Herbut12,Chamon12}, of graphene-like two-dimensional electronic systems. The masses within the effective low-energy Dirac Hamiltonian are permitted by considering spin, valley, sublattice, superconducting degrees of freedom.

The fascinating physical properties of edge states originate from the topology of  the bulk system, which is usually gapped and insulating. Solitons in Polyacetylene possess fractional charge.\cite{Su79,Jackiw12,Semenoff06} Majorana Fermions in $p_x\pm i p_y$ pairing superconductors satisfy non-Abelian fractional statistics, and can be used for topological quantum computing.\cite{Nayak,Ivanov01,Beenakker} These edge states have novel quantum transport\cite{Kane05,Setare11,Alicea12,Nomura07}, and could be induced through order parameters, proximity effects and external magnetic field.\cite{Alicea12,Leijnse12,Kohmoto07,lewenstein2007ultracold} These edge states could thus be experimentally observable.\cite{Leijnse12,Roy10} Therefore, we focus on classifying these topologically protected states at the boundary instead of topological bulk phases.

The zero-energy mode can be simply described by Dirac equation,\cite{Jackiw12, shen2013topological} 
\begin{equation}
H=\alpha p+\beta m(x)
\end{equation}
where $p$ and $m(x)$ denote the momentum and the
spatially inhomogenous mass terms, respectively. The momentum often arises in the effective theories of low-dimensional systems such as the linearization of the Fermi surface.  The mass, as order parameters, is generated by spontaneous symmetry breaking. Alternatively, the mass could be induced directly by adding external magnetic field, or the proximity of superconductors.\cite{ Ryu09,Odagiri12,Chamon12}And $\alpha,\beta$ are $8\times8$ matrices which encode the sublattice, spin and Nambu (superconducting) inner subspaces and satisfy the anticommutation relations $\{\alpha,\beta\}=0,\alpha^2=\beta^2=1$.  Solving the Dirac equation leads to the massive relativistic dispersion $\varepsilon=\pm\sqrt{p^2+m^2(x)} $. To eliminate twofold or fourfold degeneracy, the mechanism of mass competition is introduced in the Dirac Hamiltonian. By adding two or more competing masses, we achieve the isolated zero-energy mode by freezing out the redundant degrees of freedom.

\section{Zero-energy modes}
Before classifying all possible zero-energy modes, we need to clarify the relation between the mass terms and the momentum terms in Dirac equations. By combining three $2\times2$ subspaces, we classify all zero-energy modes mathematically in the $8\times8$ inner space and its three $4\times4$ subspaces, and then classify three fundamental zero-energy modes in three $2\times2$ subspaces respectively.

\begin{table}[b]
\caption{\label{tab:table1}
Six configurations of one-dimensional Dirac Hamiltonians $H=\alpha p+\beta x$ with zero-energy bound states $\psi_{0}=\chi e^{-x^2/2}$ in $2\times2$ subspaces. Three fundamental zero-energy modes that are solitons, Majorana zero modes and magnetic zero-energy modes are described by these Dirac Hamiltonians in the sublattice, Nambu, and spin space respectively.
}
\begin{ruledtabular}
\begin{tabular}{ccc}
$\alpha$&$\beta$&
$\chi$\\
\colrule
$\sigma_1$ & $\sigma_2 $& $\left[0,1\right]^T$\\
$\sigma_2$ & $\sigma_1 $& $\left[1,0\right]^T$\\
$\sigma_2$ & $\sigma_3 $& $\frac{1}{\sqrt{2}}\left[1,-1\right]^T$\\
$\sigma_3$ & $\sigma_2 $& $\frac{1}{\sqrt{2}}\left[1,1\right]^T$\\
$\sigma_3$ & $\sigma_1 $& $\frac{1}{\sqrt{2}}\left[-1,i\right]^T$\\
$\sigma_1$ & $\sigma_3 $& $\frac{1}{\sqrt{2}}\left[1,i\right]^T$\\
\end{tabular}
\end{ruledtabular}
\end{table}

Firstly, we clarify the relation between the masses and the momentums by an example, the Su-Schrieffer-Heeger (SSH) model of Polyacetylene\cite{Su79,Su80}, representing spinless electrons hopping on a one-dimensional lattice with staggered hopping amplitudes $|t\pm\delta t|$. At half-filling $\mu=0$, the SSH model describes a band insulator with minimal energy gap $2|\delta t|$. At midgap, the effective low-energy continuum Dirac Hamiltonian is described by,
\begin{equation}
H=p \sigma_1+m(x) \sigma_2
\end{equation}
with $m(x)=2\delta t$ and $ \sigma_1,\sigma_2$ are Pauli matrices. The bulk winding number is either $\nu=1$ for A phase $\delta t<0$ or $\nu=0$ for B phase $\delta t>0$.\cite{shen2013topological,Xiao10} The soliton emerges at the domain wall between A phase and B phase, which is characterized by the topological invariant $\Delta\nu=|\nu_A-\nu_B |=1$; and this is the bulk-boundary correspondence.\cite{Jackiw12,Ryu10} To solve the zero-energy Schršdinger equation $H\psi_0=0$ explicitly, we get the boundstate ansatz for the soliton,
\begin{equation}
\psi_0=\chi e^{-\int_0^x m(x')dx'}
\end{equation}
where $\chi=\left[0,1\right]^T$ as shown in Table~\ref{tab:table1}. The boundstate $\psi_0$ that decay exponentially into the bulks is localized at the domain wall. The mass is well defined in bulk phases, but disordered at the domain walls. However, different from the hyperbolic tangent function mass forms,\cite{Su80} we linearize the mass term at domain wall and link consistently to the bulk masses. The bulk Dirac Hamiltonian remains gapped and its topological property is still unchanged, thus we introduce the linear mass term,
\begin{equation}
m(x)=\xi x
\end{equation}
where $\xi=\frac{\partial m(x)}{\partial x}|_{x=0}$ denotes the spatial derivative of the mass term at the domain wall. The linear mass is restrained at domain wall and disappears linearly when the energy gap closes. For simply, we could reduce the mass per unit length $\xi=1$ and rewrite the zero-energy modes Dirac Hamiltonian,
\begin{equation}
H=\alpha p+\beta x
\end{equation}
where $|x|<L_{DW}$ and $L_{DW}$ is the width of the domain wall (or the boundary).

There is a dual relation in Dirac Hamiltonian between the linear mass and the momentum of zero-energy modes. Since both the momentum and mass term are linear, we replace $p\rightarrow\frac{\partial x}{i}$ for $H=\alpha p+\beta x$ in x-space (real space) and $x\rightarrow -\frac{\partial p}{i}$ for $\bar{H} =\beta p-\alpha x$ in p-space (momentum space), and solve the zero-energy states respectively,
\begin{eqnarray*}
H &=-i\alpha \partial x+\beta x, \quad \psi_0(x)= \chi e^{-x^2/2}\\
\bar{H} &=-i \alpha \partial p+\beta p,\quad \psi_0(p)= \chi e^{-p^2/2}
\end{eqnarray*}
In the zero-energy mode Dirac Hamiltonian, the mass-momentum duality states that the linear mass term in x-space  is equivalent to act as Òthe momentum termÓ in p-space. In the dual spaces (x-space and p-space) , the two Hamiltonians $(H,\bar{H})$ have the same momentum matrix$\alpha$ and same mass matrix $\beta$. The zero-energy mode $\psi_0$ in the representation of x-space and p-space are both Gaussian localized and self-similar. In p-space, the momentum term $\alpha p$ is similar as Òthe linear mass termÓ and the mass term $\beta x$ as Òthe momentum termÓ; and there seems to exist a Òdomain wallÓ in p-space. Since the two Hamiltonians $(H,\bar{H})$ describe the same zero-energy mode, thus the mass-momentum duality helps us simplify our mass classification of zero-energy modes in x-space. In Table~\ref{tab:table1}, we classify mathematically six nondegenerate zero-energy modes in the $2\times2$ subspace for one-dimensional Dirac Hamiltonian $H=\alpha p+\beta x$, where the matrices $\alpha$ and $\beta$ are Pauli matrices.

In general, to represent Dirac Hamiltonians in $8\times8$ spanned inner space\cite{Ryu09,Herbut12,Herbut11}, we define the 64 8-dimensional matrices set $X_{\tau\mu\nu}=\sigma_{\tau}\otimes\sigma_{\mu}\otimes\sigma_{\nu}$including the identity matrix $X_{000}=\sigma_{0}\otimes\sigma_{0}\otimes\sigma_{0}$, where the indices $\tau,\mu,\nu$  denote Nambu, spin, and sublattice channel. In $8\times8$ inner space, there are $2016(63\times32)$ kinds of configurations for $\{\alpha, \beta\}=0$, and the zero-energy modes are fourfold. In three $4\times4$ subspaces, there are $120(15\times8)$ kinds of configurations respectively, and the zero-energy modes are twofold, which are included in the configurations of the $8\times8$ inner space by spanned the rest subspace. All mass and momentum matrices could be classified in terms of the three following symmetries\cite{ Schnyder08,Chamon12}: time-reversal symmetry (TRS) $X_{021} H^* (-p) X_{021}=H(p)$, sublattice symmetry (SLS) $X_{003} H(p) X_{003}=-H(p)$, and particle-hole symmetry (PHS) $X_{100} H^T (p) X_{100}=-H(p)$; and then we find that many configurations belong to same symmetry classes physically.

Secondly, there are three fundamental zero-energy modes that are solitons, Majorana zero modes, and magnetic solitons in one-dimensional Dirac systems lying on the sublattice, Nambu and spin subspaces respectively. In sublattice subspace, we write a general spinless tight-binding Hamiltonian,
\begin{eqnarray}
H  &=&-\sum_{i}\mu_{A}c_{A,i}^{\dag}c_{A,i}-\sum_{i}\mu_{B}c_{B,i}^{\dag}c_{B,i}\notag\\
	&+&\sum_{i}(t_1c_{A,i}^{\dag}c_{B,i}+t_2c_{A,i}^{\dag}c_{B,i-1}+h.c.)
\end{eqnarray}
where $\mu_{A},\mu_{B}$ are chemical potentials of two sublattices A and B, $t_1,t_2$ label the inter-sublattice hopping and intra-sublattice hopping. The first limit is the SSH model that corresponds to $t_{1,2}=t\pm\delta t$ and $\mu_{A}=\mu_{B}=0$, where $\delta t$ is the staggered hopping. In momentum space we configure the SSH Dirac Hamiltonian by linearizing the energy dispersion as the momentum term (expand around $k_0=\pi$), and then the staggered hopping acts as the mass term.\cite{Su79,Jackiw12} The second limit is the staggered chemical potential $\mu(x)\sigma_3$ as mass term that corresponds to $\mu_{A}=-\mu_{B}=\mu(x)$, and then the momentum term is supported by two choices of the hopping, which are $t_{1,2}=t$ and $t_{1,2}=\pm\delta t$. In summary, there are three configuration of zero-energy modes Dirac Hamiltonian in sublattice space, as showed in Table~\ref{tab:table1},which are (i) linearization at Fermi level as momentum term $\sigma_1$ and staggered hopping as mass term $\sigma_2$; (ii) linearization at Fermi level as momentum term $\sigma_1$ and staggered chemical potential as mass term $\sigma_3$; (iii) staggered hopping as momentum term $\sigma_2$ and staggered chemical potential as mass term $\sigma_3$. The last two configurations are different from the SSH model in sublattice subspace.

In Nambu subspace, the Kitaev toy model Hamiltonian describing a one-dimensional p-wave spinless superconductivity reads\cite{Kitaev01,Thakurathi},
\begin{eqnarray}
H  &=&-\sum_{i}\mu c_{i}^{\dag}c_{i}-\frac{1}{2}\sum_{i}(t c_{i}^{\dag}c_{i+1}+\Delta c_{i}c_{i+1}+h.c.)
\end{eqnarray}
Upon transforming to the momentum space, the Hamiltonian is written by,
\begin{equation}
H=-\varepsilon_k \sigma_3+\left[Re(\Delta)\sigma_2+Im(\Delta)\sigma_1\right]\sin k
\end{equation}
with the normal metal energy $\varepsilon_k=\nu-t \cos k$ and the p-wave pairing potential $\Delta$. In this toy model, the normal metal order $\varepsilon_k \sigma_3$ acts as the mass term and the pairing potential as the momentum term, as showed in Table~\ref{tab:table1}. The momentum term has two parts $Re(\Delta)\sigma_2$ and $Im(\Delta)\sigma_1$ due to the arbitrary phase of pairing potential, and by the gauge transformation, we could choose one of the two parts of $\Delta$. To see Majorana zero modes explicitly, we decompose the electron annihilation operator into two Majorana fermions via $ c_{i}=\frac{1}{2}(\gamma_{B,i}+i\gamma_{A,i})$, which obey the canonical Majorana fermion relations. In Majorana representation with $\mu=0$, Hamiltonian becomes
\begin{equation}
H  =-\frac{i}{4}\sum_{i}\left[(\Delta+t)\gamma_{B,i}\gamma_{A,i+1}+(\Delta-t)\gamma_{A,i}\gamma_{B,i+1}\right]
\end{equation}
which is also called the modified SSH model because of the staggered hopping $\Delta \pm t$.\cite{shen2013topological} When $\Delta = t\neq 0$, the topological phase appears and two coupled Majorana zero modes emerge at adjacent lattice ends, which are distinguished by the $Z_2$ topological invariant $\nu=-1$ (while $\nu=1$ in trivial phase).\cite{Alicea12}

In spin subspace, the projections of spin-orbit coupling $p\cdot \sigma$ act as momentums and that of the external magnetic field $\vec{B}\cdot \sigma$ as masses, as showed in Table~\ref{tab:table1}. For a Ferromagnet with spin-orbit coupling in momentum space, one yields\cite{shen2013topological}
\begin{equation}
H=\lambda \sin k \sigma_1+(M-4g\sin^2 \frac{k}{2})\sigma_3
\end{equation}
where $\lambda$ is the strength of spin-orbit coupling $\sigma_1$ and $M$ is the Zeeman energy. The extra term $-4g\sin^2 \frac{k}{2}$ avoids the fermion-doubling problem.\cite{shen2013topological} We could obtain the zero-energy mode at the boundary where the Zeeman field $M(x)$ disappears. Moreover five other configurations correspond to the projections of spin-orbit coupling and external magnetic field, which could be introduced as momentums and masses to eliminate KramerÕs degeneracy of the spin degree of freedom in experiments\cite{Alicea12}.


\section{Mass manipulation}

\begin{table*}
\caption{\label{tab:table2}Mass manipulation and experimental implementation of zero-energy modes in one-dimensional Dirac systems. All masses and momentums are classified by their subspaces in this table. Some masses as order parameters could be induced by spontaneous symmetry breaking, and some masses could be induced by adding external field and the proximity effect. Within our mass classification scheme, we could manipulate the masses and momentums to realize the isolated zero-energy modes in experiments.}
\begin{ruledtabular}
\begin{tabular}{lll}
\textbf{subspaces}&\textbf{masses}&\textbf{momentums}\\ \hline
sublattice				&staggered chemical potential;     		   &linearization at Fermi level; \\
	         				&staggered hopping; CDWs.           		   &staggered hopping$(k=0)$;CDWs.\\
spin           				&external magnetic field; SDWs.  		   &spin-orbit coupling;SDWs.\\
Nambu				&normal metallic states.  				   &p-wave pairing SC.\\
sublattice+spin			&staggered magnetic field; SDWs.   	     	   &SDWs.\\
Nambu+spin			&s-wave pairing SC;etc.  	   			   &d-wave pairing SC(expand around gap nodes).\\
sublattice+Nambu		&p-wave pairing SC$(k\neq0)$.  		   &p-wave pairing SC.\\
sublattice+Nambu+spin	&s-extended pairing SC;	   			  &d-wave pairing SC(expand around gap nodes).\\
	         				&staggered s-wave pairing SC$(S^{\pm})$;etc. &		\\
\end{tabular}
\end{ruledtabular}
\end{table*}
To achieve isolated zero-energy modes in $4\times4$ subspaces, we must introduce the competing mass terms into one-dimensional Dirac systems. Based on the competing, we could manipulate the masses and momentums to implement the applications of the isolated zero-energy modes. To eliminate the twofold degeneracy in $4\times4$ subspaces, e.g. the combined spin and Nambu subspaces, we put two mass terms into the Dirac Hamiltonian,
\begin{equation}
H=\alpha p+\beta_1 m_1+\beta_2 m_2
\end{equation}
where $\alpha, \beta_{1,2}\in X_{\mu\nu},\{\alpha, \beta_{1,2}\}=0$ and the masses $m_{1,2}=m_{1,2} (x)$. Two cases are classified by the two masses relation: $180(15\times4\times3)$ choices of mass competition satisfying $\left[\beta_{1},\beta_{2}\right]=0$ and $240(15\times4\times4)$ choices of mass cooperation satisfying $\{\beta_{1},\beta_{2}\}=0$.

For mass competition satisfying $\{\alpha, \beta_{1,2}\}=0 $, $\left[\beta_{1},\beta_{2}\right]=0$, we square the  Hamiltonian $H^2=p^2+m_1^2+m_2^2+2m_1 m_2 \beta_{1}\beta_{2}$, and one obtains the energy dispersion with $ \beta_{1}^2\beta_{2}^2=1$,
\begin{equation}
H=\pm\sqrt{p^2+(m_1\pm m_2)^2}
\end{equation}
where the energy dispersion splits into four bands when $m_1\neq\pm m_2$; and we define two gaps,
\begin{eqnarray}
\delta m &=&\min{|m_1\pm m_2|}\\
\overline{m} &=&\max{|m_1\pm m_2|}
\end{eqnarray}
Then the bands with mass gap $\overline{m}$ in the limit $\delta m\ll\overline{m}$ could be projected away, and thus one obtains the effective low-energy bands $H_{eff}=\pm\sqrt{p^2+(\delta m)^2}$. Correspondingly, we project the $4\times4$ subspaces into the $2\times2$ subspaces and rewrite the effective zero-energy modes Dirac Hamiltonian,
\begin{equation}
H_{eff}=\bar{\alpha} p+\bar{\beta} \delta m
\end{equation}
where $\bar{\alpha},\bar{\beta}$ are Pauli matrices. The projection into $2\times2$ subspaces eliminates KramerÕs degeneracy; and the isolated zero-energy modes could be described in Dirac Hamiltonian $H_{eff}$ with effective mass $\delta m$, as shown in Table~\ref{tab:table1}. The effective mass switches the energy bands into topological phases, and the zero-energy mode appears at the criticality of quantum phase transitions.\cite{Alicea12}For example the spin-polarized p-wave pairing superconductor, by adding a strong magnetic field to freeze the spin degree of freedom, is described by the two masses Dirac Hamiltonian satisfying,
\begin{equation}
\{\alpha, \beta_{1}\}=0 , \{\alpha, \beta_{2}\}=0 ,\left[\beta_{1},\beta_{2}\right]=0
\end{equation}
where $\alpha=X_{20}$ is the momentum term (the real part of p-pairing $\Delta$)and $\beta_{1}=X_{30}$ is the mass term (the normal order $\varepsilon_k$), and $\beta_{2}=X_{33}$ is Zeeman field in spin subspace spanned by Nambu subspace. The two masses compete to lift the degeneracy, and the isolated Majorana zero modes is induced. Moreover, to eliminate the four-fold degeneracy in $8\times8$ subspaces, we introduce three masses obeying, $\left[\beta_{1},\beta_{2}\right]=\left[\beta_{2},\beta_{3}\right]=\left[\beta_{3},\beta_{1}\right]=0$, which commutate with each other and constitute a complete set,
\begin{equation}
H=\alpha p+\beta_1 m_1+\beta_2 m_2+\beta_3 m_3
\end{equation}
The mechanism of mass competition eliminates the four-fold degeneracy and realizes the isolated zero-energy modes in the reduced $2\times2$ subspaces. From the mass-momentum duality, the momentum competition is equivalence to the mass competition; in other words, we could manipulate the momentums as the masses to eliminate KramerÕs degeneracy in the Dirac Hamiltonian.

For mass cooperation satisfying $\{\alpha, \beta_{1,2}\}=0 $, $\{\beta_{1},\beta_{2}\}=0$, we obtain the energy dispersion by squaring the Hamiltonian directly,
\begin{equation}
H=\pm\sqrt{p^2+m_1^2+ m_2^2}
\end{equation}
We define a total mass matrix $\beta=\frac{m_1}{m}\beta_1+\frac{m_2}{m}\beta_2$, where $m=\sqrt{m_1^2+ m_2^2}$), and then rewrite the Hamiltonian in the standard form $H=\alpha p+\beta m$. Thus, the zero-energy mode is still two-fold and trivial.

There is another possible mass competition where the momentum commutates to one mass and anitcommutates to another. As Fu and Kane's pioneering proposals\cite{FuLiangKane, Alicea12}, a 2D topological insulator (2DTI) with the superconducting proximity effect and a Zeeman field, the model Hamiltonian is described by $H=H_{TI}+H_Z+H_{\Delta}$, which obeys $\{\alpha, \beta_{1}\}=0  ,\left[\beta_{1},\beta_{2}\right]=0$, but $\left[\alpha, \beta_{2}\right]=0$. Explicitly, the momentum term $\alpha=X_{33}$ of the 2DTI Hamiltonian $H_{TI}=\int dx \psi^{\dag}(-iv\partial_x \sigma_3-\mu)\psi$, the first mass term $\beta_1=X_{31}$ of the Zeeman field $H_{Z}=-h\int dx \psi^{\dag}\sigma_1\psi$ where $h\geqslant 0$ is Zeeman energy, and the second mass term $\beta_2=\frac{\mu}{\sqrt{\mu^2+\Delta^2}}X_{33}+\frac{\Delta}{\sqrt{\mu^2+\Delta^2}}X_{22}$ of the combined chemical potential($X_{33}$) of $H_{TI}$ and s-wave pairing superconductor($X_{22}$) of $H_{\Delta}=\int dx \Delta(\psi_{\uparrow}\psi_{\downarrow}+h.c.)$ , are competing together to mediate a spinless p-wave paring effectively. Actually without the mass term $\beta_2$, the Hamiltonian $H_{TI}+H_Z$ supports the magnetic zero-energy modes for the electron band in spin subspace; by mixing the hole band with the mass term $\beta_2$ that commutates both $\alpha$ and $\beta_{1}$, one obtains the effective isolated Majorana zero-energy modes at the topological criticality of $h=\sqrt{\mu^2+\Delta^2}$. 

As we discussed above, we can engineer the zero-energy modes in Dirac Hamiltonians by manipulating the masses and momentums. In the low-energy effective theory, both the momentums and masses are viewed as order parameters dynamically generated by spontaneous symmetry breaking or externally induced by strong fields. \cite{Chamon12}Order parameters originate in the channelsÕ instabilities of the effective interaction, i.e. CDW channel, SDW channel, and superconducting channel.\cite{frg} 

In Table~\ref{tab:table2}, we list all possible order parameters and external fields in one-dimensional Fermion systems within our mass classification scheme. Linearization at Fermi level or staggered hopping between sublattices, spin-orbit coupling or equivalently SDWs\cite{Ryu09}, and p-wave pairing potentials (in Nambu subspace) or d-wave paring potentials expanding around gap nodes (in Nambu+spin subspace), support the momentums in different subspaces. As well as the masses are classified in different subspaces in the table, such as staggered magnetic field and s-wave pairing superconductors.\cite{Guo11,Alicea12} Interestingly, staggered chemical potential acts as the mass term in sublattice subspace, staggered magnetic field acts as the mass term in the combined sublattice and spin subspaces, and then staggered s-wave pairing superconductor (in sublattice+Nambu+spin subspace) could act as the mass term in the combined sublattice, spin and Nambu subspaces. And even a p-wave pairing potential could also be the mass term if expand far away from $k=0$. 

Many order parameters, as widely available ingredients to mass manipulation in experiments, support the mass competing to generate the isolate zero-energy modes. For example, to realize the isolated Majorana zero modes experimentally\cite{LutchynRoman, OregYuvalandRefael, Alicea12}, we need a metallic nanowire with spin-orbit coupling, an s-wave pairing superconductor, and a modified magnetic field in Table~\ref{tab:table2}. By mass competition, we read the following parts of the Hamiltonian in $4\times4$ subspace: the spin-orbit coupling offers a momentum matrix $X_{32}$, the magnetic field offers a mass term $X_{33}$, and both the chemical potential and s-wave pairing potential offer another mass term $\tilde{X}_m=\frac{\mu}{\sqrt{\mu^2+\Delta^2}}X_{30}+\frac{\Delta}{\sqrt{\mu^2+\Delta^2}}X_{22}$. We examine the competing relation of two masses and the momentum,

\begin{eqnarray*}
&\{X_{32},  X_{33}\}=0 \\
&\{X_{32},  \tilde{X}_m\}=0\\
&\left[X_{33}, \tilde{X}_m\right]=0
\end{eqnarray*}
where two masses commutate to lift the two-fold degeneracy. The isolated Majorana zero mode emerges at the criticality of $h=\sqrt{\mu^2+\Delta^2}$ when the competing mass disappears.\cite{Alicea12}

\section{Summary}
In conclusion, we focus on zero-energy modes at the domain wall, not on topological phases in one-dimensional systems. To describe zero-energy modes, we introduce the momentum-mass duality into Dirac Hamiltonian by linearizing the mass term at the boundary. Then we find that three fundamental zero-energy modes: solitons, Majorana zero modes\cite{Neupert10,Santos11}, and magnetic zero-energy modes. By combining three subspaces, we classify zero-energy modes in higher subspaces, where zero-energy modes are fourfold degenerate. Luckily, the mass competition could eliminate Kramer's degeneracy and generate the isolated zero-energy mode. 

In this paper, we list all possible masses and momentum to implement at experiments in Table~\ref{tab:table2}. Cold-atom systems can be used to simulate the topologically protected zero-energy modes in condensed matter physics, and they offer flexible conditions of control and observation, such as mimic Hubbard, disordered and spin models.\cite{lewenstein2007ultracold} Recently, synthetic gauge fields, effective magnetic field and spin-orbit coupling have been realized in cold atomic gases, which allows one to study zero-energy modes and topological phases by using optical lattices.\cite{lin2009synthetic}As a result, various masses and momentums in Dirac Hamiltonians can be explicitly controlled experimentally and our mass classification scheme may be easily verified. The zero-energy modes Dirac Hamiltonians could be extended into 2D and 3D Fermi or Bose systems to describe topologically protected edge states, and surface states.\cite{Pereg-Barnea12,Semenoff2012}

\begin{acknowledgments}
This work was supported by 973 Program under Grant No. 2011CB922103, and by the
National Natural Science Foundation of China under Grant Nos. 60825402 and  11023002. 
\end{acknowledgments}
\bibliography{apssamp}
\end{document}